\documentclass[twocolumn,aps,pra,showpacs,amsmath,amssymb,superscriptaddress]{revtex4-1}

\usepackage[usenames]{color}
\usepackage{graphicx}
\usepackage{amsmath,stackrel}
\usepackage{soul}

\usepackage{mathrsfs}

\begin{document}

%\title{Enhanced Association and Dissociation of Feshbach Molecules from Atomic Gases Enabled in Microgravity}
\title{Enhanced association and dissociation of heteronuclear Feshbach molecules in a microgravity environment}

%\title{Microgravity enhanced association and dissociation of heteronuclear Feshbach molecules}
%\title{Feshbach molecule dynamics relevant to a microgravity environment}
%\title{Association and dissociation of Feshbach molecules in a microgravity environment}

\author{J. P. D'Incao}
\affiliation{JILA, University of Colorado and NIST, Boulder, Colorado, USA}
\affiliation{Department of Physics, University of Colorado, Boulder, Colorado, USA}
\author{M. Krutzik}
\affiliation{Humboldt-Universit\"at zu Berlin, Institut f\"ur Physik, Berlin, Germany}
\affiliation{Jet Propulsion Laboratory, California Institute of Technology, CA, USA}
\author{E. Elliott}
\author{J. R. Williams}
\affiliation{Jet Propulsion Laboratory, California Institute of Technology, CA, USA}

\begin{abstract}
We study the association and dissociation dynamics of weakly bound heteronuclear Feshbach 
molecules using transverse RF-fields for expected parameters accessible through the microgravity environment of NASA's Cold Atom 
Laboratory (CAL) aboard the International Space Station, including temperatures at or below nK and atomic densities as low as $10^8$/cm$^3$.
We show that under such conditions, thermal and loss effects can be greatly suppressed resulting in high efficiency in both association and 
dissociation of Feshbach molecules with mean size exceeding 10$^4a_0$, and allowing for the coherence in atom-molecule transitions to be 
clearly observable. Our theoretical model for heteronuclear mixtures includes thermal, loss, and density effects in a simple and conceptually clear manner. 
We derive the temperature, density and scattering length regimes of $^{41}$K-$^{87}$Rb that allow optimal association/dissociation 
efficiency with minimal heating and loss to guide future experiments with ultracold atomic gases in space.
\end{abstract}


\maketitle

\section{Introduction}

Association and dissociation of ultracold Feshbach molecules have been an enabling probe of fundamental physics 
throughout the last decade. \cite{chin2010rmp,koehler2006rmp}. Produced near Feshbach resonances where the atomic 
$s$-wave scattering length $a$ is magnetically tunable, these molecules have a large spatial extent and extremely weak 
binding energy. Feshbach molecules formed in Fermionic gases were crucial for the 
exploration of the BEC-BCS crossover physics \cite{greiner2003Nature,cubizolles2003prl,regal2004prla,jochim2003Science,
zwierlien2004prl,strecker2003prl,regal2004prlb,zwierlein2004prl,chin2005Science, zwierlein2005Nature}.
Their heteronuclear counterparts are an important ingredient for the creation of ultracold polar molecules 
\cite{ospelkaus2008NatPhys,ni2008Science,zirbel2008prl,wu2012prl,heo2012pra,tung2013pra,repp2013pra,
koppinger2014pra,wang2013pra,takekoshi2012pra,deh2010pra} and can be used to study universal 
few-body phenomena \cite{braaten2006PR,wang2013AAMOP,barontini2009prl,bloom2013prl,tung2014prl,pires2014prl,maier20a5prl}. 
Additionally, Feshbach molecules can be used as a source of entangled states 
\cite{greiner2005prl,poulsen2001pra,kherunssyan2002pra,kheruntsyan2005pra,yurovsky2003pra,
savage2007prl,kheruntsyan2005prl,zhao2007pra,davis2008pra,gneiting2008prl,gneiring2010pra}, 
or test the variation of fundamental constants with unprecedented sensitivity 
\cite{chin2006prl,chin2009njp,borschevsky2011pra,gacesa2014jms}.

Microgravity offers several fundamental advantages to the study of cold atoms that has sparked a growing interest 
\cite{Decadal,Turyshev2008,Binns2009} and high profile experimental efforts \cite{Muntinga2013PRL,Zoest2010Science,
Stern2009EPJD}. Most prominently, ultracold atoms released into microgravity enables interrogation and observation 
times orders of magnitude longer than their earthbound counterparts, even in a compact setup, laying the foundation 
for the next generation of space-based atom interferometer sensors applicable to both fundamental and applied 
physics \cite{Yu2006APB,STEQUEST,Williams2016NJP}. Secondly, the removal of a linear gravitational potential 
allows for enhanced delta-kick cooling and adiabatic decompression to conserve phase space density while lowering 
both temperature and density \cite{Chu1986,Ammann1997PRL,Myrskog2000PRA,Leanhardt2003Science}, 
opening the door to a new parameter regime of ultralow densities and ultracold temperatures.
Lastly, microgravity negates the ``gravitational sag" that gives a mass-dependent 
displacement of ultracold gases from their trap centers \cite{Davis2002,Hansen2013PRA}, limiting the
overlap of multiple, distinct atomic species prepared at low temperatures in a common trap. 
Eliminating this sag removes a dominant systematic error in equivalence principle measurements that use 
dual species atomic clouds as quantum test masses. Therefore, the unique environment of space 
provides a means to study high phase-space densities of single- or multi-species gases in new regimes of temperature and
density held by vanishingly-weak traps or even in extended free fall.

To this end, NASA's Cold Atom Laboratory (CAL) is scheduled for launch in 2017 as a multi-user facility to the 
International Space Station (ISS)
to study ultracold atoms, dual-species mixtures, and/or quantum degenerate gases of bosonic $^{87}$Rb and $^{39}$K 
or $^{41}$K in persistent microgravity \cite{Thompson2013}. CAL is designed as a simple, yet versatile, 
experimental facility that features numerous core technologies for contemporary quantum gas experiments including 
tunable magnetic fields [steady state, radio frequency (RF) and microwave] for atomic state manipulation and access 
to homonuclear or heteronuclear Feshbach resonances, Bragg beams for dual species atom interferometry, and 
high-resolution absorption imaging capabilities.

In this paper we develop a simple and intuitive description of the association and dissociation of heteronuclear Feshbach 
molecules using oscillating magnetic fields. We further apply this general treatment to $^{41}$K$^{87}$Rb molecules within 
the microgravity regime at CAL. 
Our formulation highlights the coherent properties of association and dissociation of Feshbach molecules
and qualitatively includes the effects of density, temperature, and 
few-body losses.
%Our formulation is non-perturbative and includes the effects of density, temperature, and 
%few-body losses.
 Our results are consistent with previous experiments performed at the usual temperatures and densities 
relevant for terrestrial experiments \cite{klempt2008pra,weber2008pra}. We find that the efficiency of association and dissociation 
of extremely weakly bound Feshbach molecules are greatly enhanced in the CAL environment allowing for 
observation of their coherent properties with high accuracy and minimal incoherent effects associated with heating and losses. 
From our analysis, we identify the conditions (in terms of the experimentally relevant parameters) a system needs to 
satisfy in order to achieve high efficiency for both association and dissociation.

\section{Molecular Association and Dissociation} \label{Theory}

The scheme we employ for Feshbach molecule association and dissociation uses an oscillating RF magnetic-field 
(transversal to the direction of the main Feshbach field) which couples atomic hyperfine states whose
$\Delta m_{f}=\pm1$, where $m_f$ is the azimuthal component of the 
hyperfine angular momentum $f$.
Provided that the magnetic field modulation frequency, $\omega/2\pi$, is resonant with a single hyperfine transition for one of the species (i.e., 
no other hyperfine states are nearby), the interaction that defines the coupling between the relevant states can be stated as
\begin{eqnarray}
W(t)=\frac{\hbar\Omega}{2}\Big(|\alpha\rangle\langle\alpha'|+|\alpha'\rangle\langle\alpha|\Big)\cos\omega t,\label{WCoupling}
\end{eqnarray}
where $\Omega/2\pi$ is the atomic Rabi-frequency while $|\alpha\rangle\equiv|f_\alpha m_{f_\alpha}\rangle$ 
and $|\alpha'\rangle\equiv|f_{\alpha'} m_{f_{\alpha'}}\rangle$ 
are the two hyperfine states satisfying the condition $\Delta m_{f}=\pm1$. 
One interesting aspect of this scheme is that the free-atom initial state can be stable at magnetic fields near the Feshbach resonance, avoiding large three-body losses that 
otherwise arise for resonantly interacting Bosonic gases. Relevant to CAL, an initial weakly interacting mixture of Rb and K atoms in the $|10\rangle$ and $|11\rangle$ states, 
respectively, would be available to access Feshbach molecules in the $|11\rangle$ atomic states of both species at magnetic fields near the broad resonance at 39.4 Gauss 
\cite{weber2008pra,klempt2007pra,simoni2008pra,thalhammer2009njp}.
However, we will keep our theoretical model general. 

Our model for molecular association and dissociation is derived from the Floquet formalism \cite{chu2004pr},
appropriate for time-periodic Hamiltonians, and assumes zero-range interatomic interactions \cite{huang1957pr}. Although more 
sophisticated theoretical models exist \cite{chin2010rmp,koehler2006rmp}, the use of zero-range interactions will allow us to extract
the important parameters controlling the various aspects of molecular association and
dissociation relevant for experiments. Within our framework, the Floquet Hamiltonian for two atoms in the presence of
an external field (periodic in time) is written as
\begin{eqnarray}
\mathscr{H}_F=H+|\beta\rangle W(t)\langle\beta|-i\hbar\frac{\partial}{\partial t}
\end{eqnarray}
where $H$ is the bare, time-independent, two-atom Hamiltonian whose eigenstates are $\psi_{\nu}$ with energies $E_{\nu}$, and $|\beta\rangle$ is the internal state
for the spectator atom, i.e., the atom not affected by the external field. 
We seek for the solutions of the Floquet Schr\"odinger equation, $\mathscr{H}_F\Psi_F=\varepsilon\Psi_F$,
with quasi-eigenenergy $\varepsilon$ and quasi-eigenstate
\begin{eqnarray}
\Psi_F(\vec{r},t)=\sum_{n\nu}c^n_{\nu}\psi_{\nu}(\vec{r})e^{in\omega t}.\label{PsiF}
\end{eqnarray}
In the above equation, $\vec{r}$ is the interparticle vector, and $n$ is the photon number. 
Considering only $s$-wave interactions, the bare wavefunction can be written as
\begin{eqnarray}
\psi_\nu(\vec{r})=\frac{1}{2}\sqrt{\frac{1}{\pi}}\frac{f_\nu(r)}{r}|S_\nu\rangle
\end{eqnarray}
where $|S_\nu\rangle=\{|\alpha\beta\rangle,|\alpha'\beta\rangle\}$ represents the
two-atom spin states and $f_\nu$ is their corresponding radial wave function. 
Now, using Eqs.~(\ref{PsiF}) we can write the 
Floquet Schr\"odinger equation, after projecting out the 
base $\psi_{\nu}(\vec{r})e^{in\omega t}$, as 
\begin{align}
\sum_{n'\nu'}\Big[E_{\nu}\delta_{nn'}\delta_{\nu\nu'}+\frac{\hbar\Omega^{\nu}_{\nu'}}{2}\big(\delta_{n,n'+1}+\delta_{n+1,n'}\big)\nonumber\\
+(n\hbar\omega-\varepsilon)\delta_{nn'}\delta_{\nu\nu'}\Big]c^{n'}_{\nu'}=0,\label{EigenEq}
\end{align}
where 
\begin{eqnarray}
\Omega^{\nu}_{\nu'}=\Omega\int_0^\infty f^{\ast}_{\nu}(r)f_{\nu'}(r)dr,\label{Lambda}
\end{eqnarray}
defines the two-atom Rabi-frequency. 
Note that $\Omega^{\nu}_{\nu'}$ is non-zero only for values of $\nu\neq\nu'$ satisfying the selection rules ($\Delta m_{f}=\pm1$) 
imposed by the form of the atom-external field coupling in Eq.~(\ref{WCoupling}). 
The solutions of Eq.~(\ref{EigenEq}) fully determine the time-evolution of the atomic and molecular states coupled by the external field. 
In practice, for values of $\hbar\Omega^{\nu}_{\nu'}$ much smaller than any other energy scale in 
the problem, only states with $|n|=0$ and $1$ are necessary to accurately describe the system.

\begin{figure}[hbtp]
\includegraphics[width=3.4in]{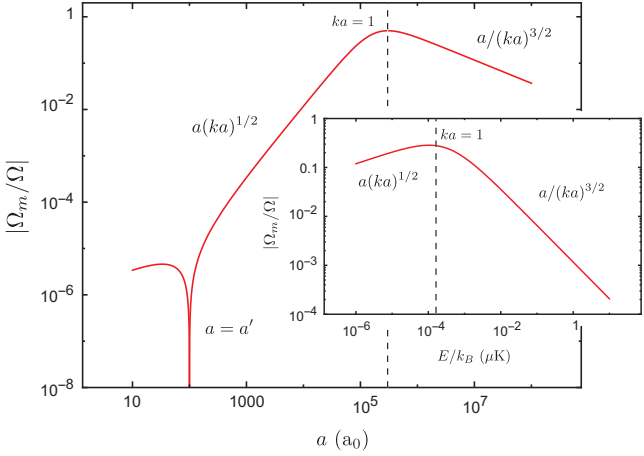}
\caption{Molecular Rabi-frequency $\Omega_m/2\pi$ (in units of the atomic Rabi-frequency, $\Omega/2\pi$) as a function of scattering length.
For this calculation we used $n_{\rm Rb}=n_{\rm K}=10^8/{\rm cm^3}$, 
and $a'=100a_0$ and $E/k_B=100$pK. On the inset we show the energy dependence of $\Omega_m$ assuming
$a=10^5a_0$.}
\label{RabiFrequency}
\end{figure}

For our present study, atoms in the spin state $|\alpha\beta\rangle$ are unbound while atoms in the $|\alpha'\beta\rangle$ state are bound in the Feshbach molecule. (Note that we will denote the corresponding states for atoms in spins $|\alpha\beta\rangle$ and $|\alpha'\beta\rangle$ as $\nu\equiv K$ and $\nu\equiv m$, respectively.) In that case, the two-atom Rabi-frequency (\ref{Lambda})
is determined from the wave functions
\begin{eqnarray}
f_{K}(r)&=&\sqrt{\frac{2\mu\epsilon_{r}}{\pi\hbar^2k}}\sin(kr-ka'),\label{fk}\\
f_{m}(r)&=&\sqrt{\frac{2}{a}}e^{-r/a},\label{fm}
\end{eqnarray}
where $\mu$ is the two-body reduced mass, $k^2=2\mu E/\hbar^2$ ($E$ is the collision energy), and
$a$ and $a'$ are the scattering lengths for atoms in the $|\alpha'\beta\rangle$ and $|\alpha\beta\rangle$ 
spin states, respectively. 
Note that in Eq.~(\ref{fk}) we have introduced an arbitrary energy scale, $\epsilon_{r}$, 
that makes both Eqs.~(\ref{fk}) and (\ref{fm}) to have the same units.
We set $\epsilon_r$ to be given in terms of the Fermi energy $\epsilon_i=\hbar^2(6\pi^2n_i)^{2/3}/2m_i$, where $n_i$
and $m_i$ are the density and mass for the atomic species $i$. 
Here we will define $\epsilon_{r}$, for 
simplicity, as $\epsilon_r=\epsilon_{\rm Rb}+\epsilon_{\rm K}$. 
Choosing $\epsilon_r$ to incorporate density effects allows our model to encapsulate 
 {\em local} properties of the system. (Similar ways to 
qualitatively account for density effects have been successfully used in few-body models
\cite{borca2003njp,goral2004jpb,stecher2007prl,sykes2014pra,corson2015pra} in order to explain molecular formation and other important properties relevant for ultracold 
gases experiments.)
As a result, substituting Eqs.~(\ref{fk}) and (\ref{fm}) in Eq.~(\ref{Lambda}), we obtain the molecular 
Rabi-frequency $\Omega_m\equiv\Omega^{K}_{m}$,
\begin{eqnarray}
\Omega_{m}(k)&=&\Omega\sqrt{\frac{4\mu\epsilon_r}{\pi\hbar^2}}\frac{(a-a')}{(1+k^2a^2)}  (ka)^{1/2}, \label{Omegam}
\end{eqnarray}
which is also density dependent ($\Omega_{m}\sim n^{1/3}$), a dependence introduced via $\epsilon_r$. 
In Fig. \ref{RabiFrequency} we show $\Omega_m$ as a function of both scattering length and energy (inset),
and indicate the low- and high-energy behavior, i.e., $ka\ll1$ [$\Omega_m\propto a(ka)^{1/2}$] and $ka\gg1$ 
[$\Omega_m\propto a/(ka)^{3/2}$], respectively.

We now have defined all elements necessary to solve Eq.~(\ref{EigenEq}).
As mentioned above, in the regime of small $\hbar\Omega_m$, we need only to consider states with $|n|=0$ and $1$.
Therefore, including only the states $\{\nu,n\}=\{K,0\}$ and $\{m,-1\}$, the eigenvalue equation 
(\ref{EigenEq}) reduces to
\begin{eqnarray}
\left(\begin{array}{cc}
E_{K} & \frac{\hbar\Omega_m}{2} \\ 
\frac{\hbar\Omega_m}{2} & E_{m}-\hbar\omega
\end{array}\right)
\left(\begin{array}{c}
c_k^{0} \\
c_m^{-1}
\end{array}\right)
=
\varepsilon
\left(\begin{array}{c}
c_k^{0} \\
c_m^{-1}
\end{array}\right),\label{TwoLevel}
\end{eqnarray}
which is formally equivalent to a two-level system in the presence of an external field within the Rotating Wave 
Approximation (RWA), whose solutions are well known \cite{CohenTannoudji1992}. As we will see next, the fact that these levels now represent
a bound molecular state and two-atom continuum state, makes it important to include thermal and loss effects 
in order to determine the time evolution process leading to association and dissociation of weakly bound molecules.

\begin{figure}[hbtp]
\includegraphics[width=3.3in]{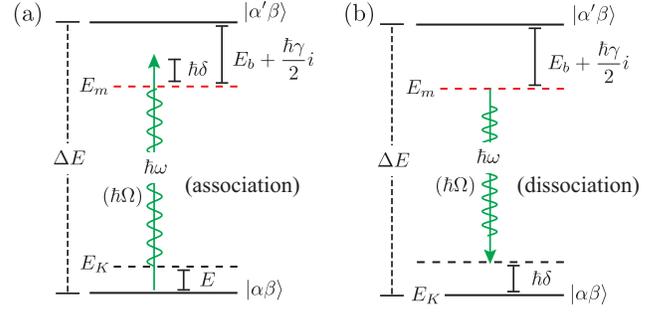}
\caption{Schematic representation for our present level scheme for (a) molecular association and (b) molecular dissociation.
Here, $\Delta E$ is the energy difference between the relevant two-atom thresholds, $|\alpha\beta\rangle$ and $|\alpha'\beta\rangle$,
$E_{b}=\hbar^2/2\mu a^2$ is the molecular binding energy, $\gamma$ is the molecular lifetime (see text) and $E$ is the energy
of the two atoms in the $|\alpha\beta\rangle$ state. Atomic and molecular states are coupled via an external field with frequency 
$\omega/2\pi$ (with detuning $\delta$) and Rabi-frequency $\Omega/2\pi$. $E_K$ and $E_m$ are given in 
Eq.~(\ref{TwoLevel}).}
\label{Scheme}
\end{figure}

The molecular association scheme, which couples atomic and molecular states with different $m_f$, is illustrated in Fig.~\ref{Scheme}~(a). 
Here, $E_K=E$ and $E_m=\Delta E-E_b-i\hbar\gamma/2$, 
where $E_b=\hbar^2/2\mu a^2$ is the binding energy of the molecular state and $\hbar\gamma$ is its corresponding width, 
introduced here to account for the finite lifetime of the molecular state due to collisions with 
other atoms and molecules. This model is valid for times shorter than $1/\gamma$. 
Assuming that at $t=0$ the atoms are 
unbound ($\nu=K$), one can show that the probability to find the atoms in the molecular state ($\nu=m$) at 
later times, $t=\tau$, is given by
\begin{eqnarray}
P_m(E,\tau)&=&e^{-\gamma \tau/2}\left(\frac{\Omega_m}{\Omega^{m}_{\rm eff}}\right)^2
\Big|\sin\left(e^{i\theta_{m}}\frac{\Omega^{m}_{\rm eff}\tau}{2}\right)\Big|^2,\label{Pm}
\end{eqnarray}
where 
\begin{eqnarray}
\Omega^{m}_{\rm eff}&=&[{\gamma^2\left(\delta+{\scriptstyle\frac{E}{\hbar}}\right)^2+(\Omega_m^2+\left(\delta+{\scriptstyle\frac{E}{\hbar}}\right)^2-{\scriptstyle\frac{\gamma^2}{4}})^2}]^{\frac{1}{4}},\label{OmegaEff}\\
\theta_m&=&\frac{1}{2}{\rm tan^{-1}}[\frac{(\delta+{\scriptstyle\frac{E}{\hbar}})\gamma}{\Omega_m^2+(\delta+{\scriptstyle\frac{E}{\hbar}})^2-{\scriptstyle\frac{\gamma^2}{4}}}].\label{ThetaM}
\end{eqnarray} 
Here, $\hbar\delta$ is the energy detuning from the molecular transition in Fig.~\ref{Scheme} (a).
Note that even for $\delta=0$ ---when one would expect the system to be on resonance--- finite energy and molecular decay 
effects can lead to an effective detuning through Eqs.~(\ref{OmegaEff}) and (\ref{ThetaM}).
Note also that $dP_m/d\tau$ in the limit $\tau\rightarrow0$ is related to the transition rate derived 
in Ref.~\cite{chin2005pra} based on the Fermi's Golden rule.
It is important to emphasize here that for the process of molecular association, since there exist a thermal 
distribution of initial states \cite{hanna2007pra}, the transition probability needs to be thermally averaged accordingly to
\begin{eqnarray}
\langle P_m(T,\tau)\rangle=\frac{2}{\pi^{\frac{1}{2}}}\int_{0}^{\infty}\frac{P_{m}(E,\tau)}{(k_BT)^{\frac{3}{2}}}E^{\frac{1}{2}}e^{-\frac{E}{k_BT}}dE.\label{PmT}
\end{eqnarray}
Here we will define the fraction of molecules formed (assuming an equal number of initial atoms of 
different species), after a square-pulse of duration $\tau$ to be given simply by 
\begin{eqnarray}
\frac{N_{m}}{N_a}=\langle P_m(T,\tau)\rangle.\label{NmNa}
\end{eqnarray}

For molecular dissociation, our scheme is represented in Fig.~\ref{Scheme}(b), leading
us to set $E_K=0$ and $E_m=E+\Delta E-E_b-i\hbar\gamma/2$ in Eq.~(\ref{TwoLevel}). 
Therefore, similarly to association, we now consider the solutions of Eq. (10) and assume 
that the system is found in the molecular state ($\nu=m$) at $t=0$. The probability 
to find the system in the unbound state ($\nu=K$) at later times, $\tau$, is 
\begin{eqnarray}
P_K(\delta,\tau)&=&e^{-\gamma \tau/2}\left(\frac{\Omega_m}{\Omega^{K}_{\rm eff}}\right)^2
\Big|\sin\left(e^{i\theta_{K}}\frac{\Omega^{K}_{\rm eff}\tau}{2}\right)\Big|^2,\label{PK}
\end{eqnarray}
where 
\begin{eqnarray}
\Omega^{K}_{\rm eff}&=&[{\gamma^2\delta^2+(\Omega_m^2+\delta^2-{\scriptstyle\frac{\gamma^2}{4}})^2}]^{\frac{1}{4}},\\
\theta_K&=&\frac{1}{2}{\rm tan^{-1}}[\frac{\delta\gamma}{\Omega_m^2+\delta^2-{\scriptstyle\frac{\gamma^2}{4}}}].
\end{eqnarray} 
Here, we note that the energy of the dissociated atoms is given by the energy detuning $\hbar\delta$
[see Fig.~\ref{Scheme} (b)]. As a result, for dissociation the $k$ dependence of $\Omega_m$ in Eq.~(\ref{Omegam}) 
needs replaced by the wavenumber associated to the energy detuning, $k_\delta^2=2\mu\delta/\hbar$, i.e.,  the relevant
Rabi-frequency is now dependent of the detuning, $\Omega_{m}\equiv\Omega_{m}(\delta)$.
We also note that, for molecular dissociation, thermal effects can only be 
introduced via the Doppler effect, i.e., molecules with different velocities will experience a different external 
field frequency, $\omega/2\pi$. However, the fact that we assume low temperatures and low frequency transitions effectively 
negates the effects of Doppler-broadening in dissociation
(see Section \ref{SecDissociation}). 
In that case, the fraction of atoms formed after a square-pulse of duration $\tau$ is given simply by 
\begin{eqnarray}
\frac{N_{a}}{N_m}=P_{K}(\delta,\tau).\label{NaNm}
\end{eqnarray}

Among the conditions for the validity of the above approach, the requirement that the system is found in 
the dilute regime, i.e., $na^3\ll1$ and $na'^3\ll1$, is of crucial importance. If such conditions are not 
satisfied nontrivial finite density effects have to be considered which are beyond the capability of 
our current model. Our model also requires $\hbar\Omega_m/E_{b}\ll1$ in order to avoid free-to-free 
transitions during both association and dissociation as well as multi-photon effects. Although our model
could be extended in order to properly include such effects, it is of experimental interest to restrict to
parameters in which $\hbar\Omega_m/E_{b}\ll1$ since this is the regime in which one can associate
or dissociate Feshbach molecules more efficiently and without generating significant heating. 

\begin{figure*}[htbp]
\includegraphics[width=6.8in]{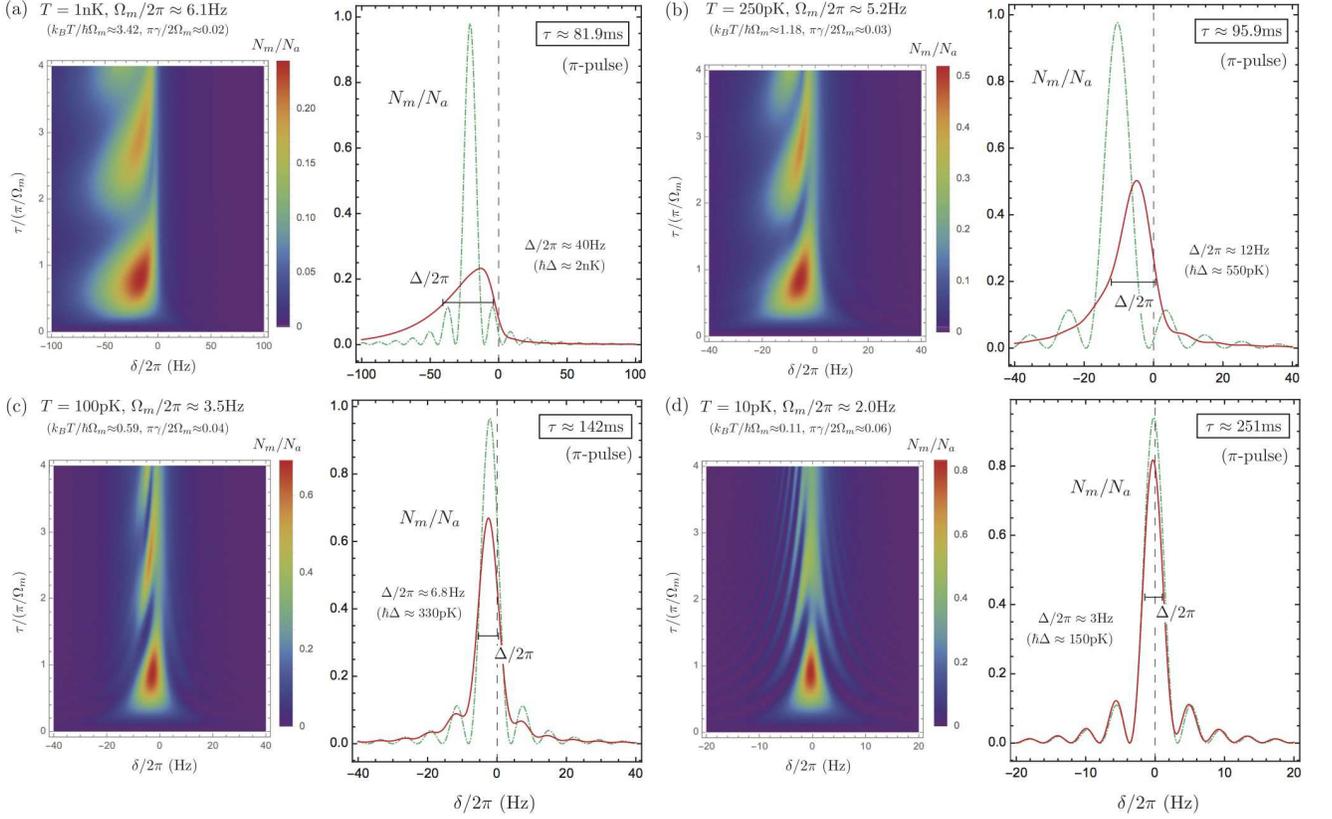}
\caption{Molecular association efficiency [Eq.~(\ref{NmNa})] for $a=10^4a_0$ ($E_b/h=642.94$Hz), $n=10^{8}$/cm$^3$, $\Omega/2\pi=0.2$kHz, and different values of $T$ 
and, consequently, $\Omega_m$, as indicated in the figure.  
For each panel (a)-(d) we display a density plot showing the pulse length, $\tau$, dependency of the molecular fraction as
a function of the detuning, $\delta$, and a figure for a $\pi$-pulse ($\tau=\pi/\Omega_m$), displaying both the thermally averaged 
results [red-solid curves given by Eq.~(\ref{PmT})] and the non-averaged results [dashed-green curves given by Eq.~(\ref{Pm}), setting $E=k_{B}T$].
In the figure we indicate the values for the ratio $k_BT/\hbar\Omega_m$ characterizing the thermal regime as well as the association linewidth, 
$\Delta/2\pi$, which ultimately sets the temperature of cloud after the pulse. The values for $T/T_c$ ($T_c\approx3.31\hbar^2n^{2/3}/m$) 
in the panels above are (a) $T/T_{c,{\rm Rb}}\approx25.3$ and $T/T_{c,{\rm K}}\approx11.4$, (b) $T/T_{c,{\rm Rb}}\approx6.33$ and 
$T/T_{c,{\rm K}}\approx2.84$, (c) $T/T_{c,{\rm Rb}}\approx2.53$ and $T/T_{c,{\rm K}}\approx1.14$,
and (d) $T/T_{c,{\rm Rb}}\approx0.25$ and $T/T_{c,{\rm K}}\approx0.11$. The validity of the model as the system is cooled into the quantum-degenerate regime ($T/T_c < 
1$) is discussed in Section \ref{SecAssociation}.
}
 \label{RF_Association}
\end{figure*}

\section{Results and Discussion} \label{SecResults}

The major focus of this study is to explore association and dissociation of Feshbach molecules in the
parameter regime relevant for CAL, i.e., we consider temperatures at or below 1nK and atomic densities as 
low as $n_{\rm K}=n_{\rm Rb}=10^8/{\rm cm^3}$. We will show that this low-temperature and low-density regime 
makes it possible to observe efficient association and dissociation as well as their corresponding coherent properties.
For our present studies, we consider fields which are far-detuned from the atomic transition, 
i.e., $\hbar\Omega/E_{b}\ll1$. Ensuring that $\hbar\Omega/E_{b}$ is small prevents single-atom spin-flip transitions, 
which can reduce the number of atoms in the initial state for association ---for the parameters used here for the atomic 
Rabi-frequency and detunings we estimate a $4\%$ probability for this effect. 

In the following, we study the case where an RF field is applied to a heteronuclear mixture of $^{87}$Rb and $^{41}$K 
initially in the $|10\rangle$ and $|11\rangle$ states, respectively, with $\Omega/2\pi=0.2$kHz for Rb. Molecular association 
and dissociation are thereby induced at $a=10^4a_0$ ($E_b/h=642.94$Hz), assuming $a'=100a_0$ 
for the initial atomic state. Therefore, we are assuming bosonic heteronuclear Feshbach molecules which are about 
10 times larger (and 100 times more weakly bound) than previously studied \cite{klempt2008pra,weber2008pra}. 
Here, three-body losses that can play an important rule at such large scattering lengths 
\cite{braaten2006PR,wang2013AAMOP}, will be greatly suppressed in the low-density, low temperature regimes available on CAL.
In fact, a more detailed analysis of Refs.~\cite{helfrich2010pra,dincao2004prl,wang2012prlb},
along with some of the experimental data from Refs.~\cite{barontini2009prl,bloom2013prl}, allow us to set $\gamma=500$mHz 
for this mixture, implying a molecular lifetime of about 2 seconds. This leaves plenty of time to associate and dissociate Feshbach 
molecules with minimal effects from loss.

\subsection{Molecular Association} \label{SecAssociation}

Figure~\ref{RF_Association} shows our results for molecular association efficiency [Eq.~(\ref{NmNa})] after 
a RF-pulse of duration $\tau$, for temperatures ranging from 1nK to 10pK. For each panel of 
Fig.~\ref{RF_Association} we display a density plot showing the pulse length dependency of the 
molecular fraction as a function of the detuning, $\delta$, and a plot for the corresponding result 
for a (square) $\pi$-pulse ($\tau=\pi/\Omega_m$). 
Note that, in Fig.~\ref{RF_Association}, we show both the thermally averaged results for molecular association 
efficiency (solid red curve) and the non-averaged results (dot-dashed green curve) 
in order to emphasize the importance of finite temperature effects.
As one can see, the Rabi-oscillation line shape is almost completely washed-out at high temperatures, while it is recovered in the low temperature regime. In fact, 
for temperatures of 10pK [Fig.~\ref{RF_Association}~(d)], atom-molecule coherences can be clearly seen, along with high association efficiency.

In Fig.~\ref{RF_Association}, the dimensionless quantity $k_BT/\hbar\Omega_m$, 
i.e., the ratio between thermal energy and the energy associated with molecule-photon coupling, helps to define the regimes 
in which thermal effects are important. For $k_BT/\hbar\Omega_m>1$ one would expect strong thermal effects since the atoms' 
motions are significant over the timescales for association. This behavior is clear from Fig.~\ref{RF_Association} where one can see that whenever 
$k_BT/\hbar\Omega_m>1$ [Figs.~\ref{RF_Association}(a) and (b)] the linewidth, $\Delta/2\pi$, is mainly 
determined by the temperature while for $k_BT/\hbar\Omega_m<1$ [Figs.~\ref{RF_Association}(c) and (d)] 
it is determined by the molecular Rabi frequency. In fact, for $k_BT/\hbar\Omega_m<1$, one can show from Eq.~(\ref{Pm}), neglecting loss effects, that the linewidth is
approximately given by,
\begin{eqnarray}
\frac{\Delta}{2\pi}\approx2\left(\frac{2}{\pi}\right)^{1/2}\frac{\Omega_m}{2\pi}.\label{Delta}
\end{eqnarray}
Note that $\hbar\Delta$ will ultimately set the temperature of the molecular cloud after the pulse. Therefore, besides
enabling higher efficiency for association, it is also of experimental interest to keep $ \Omega_m$ small 
so that minimal heating is introduced in the system. By doing so, however, it implies that longer $\pi$-pulses are necessary for association, which must be 
balanced with the time scales associated with losses.

One needs to combine low thermal broadening and minimal atomic losses to realistically observe efficient molecular association and atom-molecule 
coherent effects. These conditions are given by
\begin{align}
\frac{k_BT}{\hbar\Omega_m}&\approx\frac{0.54}{\alpha}\left[\frac{\mu^{3/4}a^{1/2}(k_BT)^{3/4}}{\hbar^{3/2}n^{1/3}}\right]\ll1, \label{ThermalC}\\
\frac{\pi\gamma}{2\Omega_m}&\approx\frac{0.85}{\alpha}\left[\frac{\tilde\gamma\hbar^{1/2}n^{2/3}a^{3/2}}{(k_BT)^{1/4}\mu^{1/4}}\right]\ll1, \label{LossC}
\end{align}
where we assumed $ka\ll1$ and $\hbar\Omega/E_b=\alpha$ in Eq.~(\ref{Omegam}), with $\alpha<1$ as required
for suppression of spin-flip transitions For the $^{87}$Rb-$^{41}$K system considered, $\alpha\approx0.31$, leading
to a 4$\%$ probability of loss from spin-flips. In Eq.~(\ref{LossC}) we define the loss rate as $\gamma=\tilde\gamma(\hbar n a/\mu)$ with 
$\tilde\gamma$ given in terms of the few-body physics controlling atomic and molecular losses \cite{wang2013AAMOP} 
---in our case, $\gamma=500$mHz which leads to $\tilde\gamma\approx4.2$. Note that, in the limit of low losses, 
Eq.~(\ref{LossC}) relates to the fraction of atoms remaining after a $\pi$-pulse, $\exp(-\pi\gamma/2\Omega_m)$ [see Eq.~(\ref{Pm})].
Therefore, Eqs~(\ref{ThermalC}) and (\ref{LossC}) can be used as a guide in order to understand the complex parameter regime that 
leads to the suppression of thermal effects combined with long lifetimes. In fact, based on our numerical calculations, we notice that 
systems with the same value for $k_BT/\hbar\Omega_m$ and $\pi\gamma/2\Omega_m$ share the same degree of thermal and loss effects.

The optimal set of parameters will, however, be determined from the combination of low temperatures and densities resulting on how strong thermal and 
loss effects are on these parameters. For instance, from Eq.~(\ref{ThermalC}), it is clear that thermal effects are more sensitive to temperature than density. 
[The opposite is true for loss effects from Eq.~(\ref{LossC}).].
In order to illustrate how to achieve an optimal set of parameters we start from typical values for ground-based experiments 
\cite{klempt2008pra,weber2008pra}: $T$=$100$nK, $n$=$10^{12}$/cm$^3$, 
$a$$=$800$a_0$, and $\Omega$=$50$kHz. In this case, although losses are not so drastic, $\pi\gamma/2\Omega_m\approx0.07$ ($\gamma$=$400$Hz 
\cite{helfrich2010pra,dincao2004prl,wang2012prlb}), thermal effects can be significant since $k_BT/\hbar\Omega_m\approx1.5$.
Although reducing the temperature to $1$nK strongly reduces thermal effects, $k_BT/\hbar\Omega_m\approx0.05$, losses now can be important, 
$\pi\gamma/2\Omega_m\approx0.22$, but not {\em drastically} important. If now the density is also decreased by a factor 10, both thermal and loss effects
should be suppressed ($k_BT/\hbar\Omega_m\approx0.10$ and $\pi\gamma/2\Omega_m\approx0.05$). 
We note, however, that such regime can only be achieved for these temperatures
and densities because the assumed scattering length ($a$=800$a_0$) is not so large. 
As shown in our results in Fig.~\ref{RF_Association}, and according to Eqs.~(\ref{ThermalC}) and (\ref{LossC}), 
as one assumes larger values of $a$, high efficiency can only be accomplished by reducing temperatures and densities 
drastically. Nevertheless, one particular motivation to explore the large $a$ limit is that the linewidth $\Delta/2\pi$ 
[see Eq.~(\ref{Delta})] should now be proportional to $1/a^{1/2}$, therefore, reducing the amount of heating generated by 
the pulse. For instance, for the parameters above, with $a$=800$a_0$, $T$=1nK, $n$=10$^{11}$/cm$^3$ and $k_BT/\hbar\Omega_m\approx0.10$, 
we obtain from Eq.~(\ref{Delta}) $\hbar\Delta\approx16$nK while, for the calculation in Fig.~\ref{RF_Association}~(d) 
---$a$=10$^4a_0$, $T$=10pK, $n$=10$^{8}$/cm$^3$ and $k_BT/\hbar\Omega_m\approx0.11$--- we obtain $\hbar\Delta\approx145$pK.

The effects of quantum degeneracy might also be important at such low temperatures for parameters used in our 
calculations ($a$=10$^4a_0$ and $n$=10$^{8}$/cm$^3$) given in Fig.~\ref{RF_Association}. The critical temperature for condensation 
is about $T_{c,{\rm Rb}}\approx40$pK and $T_{c,{\rm K}}\approx90$pK for $^{87}$Rb and $^{41}$K respectively 
 (see specific values for $T/T_c<1$ in the caption of Fig.~\ref{RF_Association}). 
In the context of molecular association, this means that atoms in the initial state
will have a narrower energy distribution than a simple thermal cloud. As a result, the thermal effects displayed in our 
calculations for when $T/T_c<1$ should be minimized, thus improving molecular conversion efficiency. For instance, 
the thermally averaged results [see Eq.~(\ref{PmT})] in Fig.~\ref{RF_Association}~(c) and (d),
should approach the non-averaged results [see Eq.~(\ref{Pm})] as the system enters in the quantum degenerate regime. This expected improvement of 
molecular association in the quantum degenerate regime has been verified experimentally in 
Refs.~\cite{hodby2005prl,thompson2005prl,papp2006prl} and analyzed in Refs. \cite{hanna2007pra,williams2006njp}. 
To emphasize the importance of quantum degeneracy, we can recast the results in 
Eqs.~(\ref{ThermalC}) and (\ref{LossC}) in terms of $T/T_c=T/T_{c,{\rm K}}=(m_{\rm K}/m_{\rm Rb})T/T_{c,{\rm Rb}}$, leading to
\begin{align}
\frac{k_BT}{\hbar\Omega_m}&\approx\frac{1.33}{\alpha}\left[\frac{(T/T_{c})^{3/4}(na^{3})^{1/6}}{(m_{\rm K}/\mu)^{3/4}}\right]\ll1, \label{ThermalCNew}\\
\frac{\pi\gamma}{2\Omega_m}&\approx\frac{0.63}{\alpha}\left[\frac{\tilde\gamma (na^{3})^{1/2}}{(\mu/m_{\rm K})^{1/4}(T/T_c)^{1/4}}\right]\ll1, \label{LossCNew}
\end{align}
and showing the reduction of thermal effects as the $T/T_c$ decreases, while keep loss 
effects under control due to the weaker dependence in Eq.~(\ref{LossCNew}) on $T/T_c$.
We also look for possible mean-field effects
that can lead to collisional frequency shifts. In our case, however, 
collisional frequency shifts $nU_0/h$ (where $U_0=2\pi\hbar^2a/\mu$) is about 0.1Hz and according
to the results in Fig.~\ref{RF_Association} this would lead to small effects ---that is also to be compared to
the local energy $\epsilon_r/h\approx5.9$Hz, as determined from the discussion preceding Eq.~(\ref{Omegam}).
However, a more precise analysis of mean-field effects, as well as effects of quantum degeneracy, 
is beyond the scope of our present study.

\subsection{Molecular Dissociation} \label{SecDissociation}

\begin{figure}[htbp]
\includegraphics[width=3.4in]{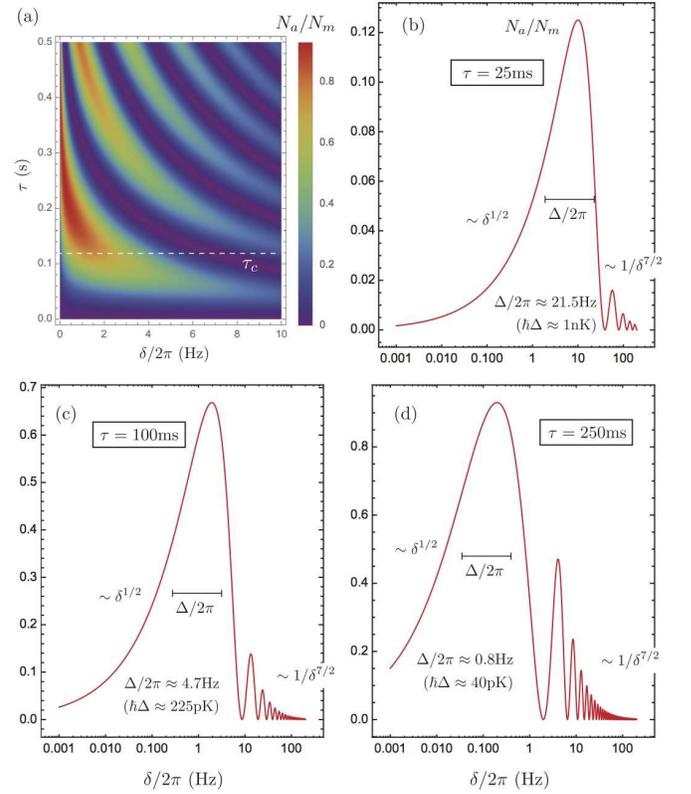}
\caption{Molecular dissociation efficiency [see Eq.~(\ref{NaNm})] as a function of the pulse duration and detuning, (a),
and for a fixed pulse duration, (b)-(c). For longer pulses we obtain high efficiency and a narrow linewidth, $\Delta/2\pi$.}
\label{RF_Dissociation}
\end{figure}

Figure~\ref{RF_Dissociation} shows some of our results for molecular dissociation efficiency [Eq.~(\ref{NaNm})],
also assuming densities of $10^8$/cm$^3$, scattering length of 10$^4a_0$ and Rabi-frequency 
$\Omega/2\pi=0.2$kHz. 
In Fig.~\ref{RF_Dissociation}(a) the density plot shows both the pulse length, $\tau$, and detuning, $\delta$, 
dependency of the fraction of atoms created after the dissociation pulse, with Rabi oscillations characterizing the 
coherent aspects of such process. In Figs.~\ref{RF_Dissociation}(b)-(c), we show the fraction of 
dissociated atoms for a fixed pulse length $\tau=$ 25ms, 100ms, and 250ms respectively. The asymmetric profile 
for of the dissociation lineshape is result
of the dependence of Rabi-frequency $\Omega_m/2\pi$ on $\delta$ [see discussion followed by Eq.~(\ref{PK})]. 
As indicated in Figs.~\ref{RF_Dissociation}(b)-(c), this dependence causes the dissociation probability in 
Eq.~(\ref{NaNm}) to vanish as $\delta^{1/2}$ for small $\delta$ and as $\delta^{7/2}$ for large $\delta$, resulting in a asymmetric lineshape.

It is important to note that while for association one expect to obtain maximum efficiency for a 
$\pi$-pulse ($\tau=\pi/\Omega_m$) at $\delta\approx0$, for dissociation (due to the dependence of $\Omega_m$ on $\delta$)
one now wants to know what is the detuning leading to maximum dissociation for a given pulse length as well as the
corresponding width of the dissociation lineshape, since that will ultimately determine the energy of the dissociated atomic pair.
As one can see from Fig.~\ref{RF_Dissociation}~(a), there is a characteristic pulse length, 
$\tau_c$, beyond which dissociation becomes efficient and the corresponding linewidth 
becomes narrow. [See horizontal dashed line in Fig.~\ref{RF_Dissociation}~(a).]
One can show that this characteristic time scale is given by
\begin{eqnarray}
\tau_c=\frac{\hbar^{5/3}\pi^{5/3}}{4[a^2(a-a')^4\epsilon_r^2\mu^3\Omega^4]^{1/3}}
\approx\frac{1.11}{\alpha^{4/3}}\left(\frac{\mu\:a^{2/3}}{\hbar\:n^{4/9}}\right),
\end{eqnarray}
where we assumed $\alpha=\hbar\Omega/E_b<1$ in order to ensure the suppression of spin-flip transitions.
(For the parameters used in our calculations in Fig.~\ref{RF_Dissociation} we obtain $\tau_c\approx93$ms.)
For long pulses, i.e., for $\tau\gg\tau_c$, the value of the detuning in which the dissociation probability is maximized 
and the corresponding linewidth are given, respectively, by
\begin{align}
\frac{\delta_{\rm max}}{2\pi}&\approx\frac{\hbar^{5}\pi^{5}}{64a^2(a-a')^4\epsilon_r^2\mu^3\Omega^4}\frac{1}{\tau^4}\nonumber\\
&\approx\frac{1.33}{\alpha^4}\frac{\mu^3 a^2}{\hbar^3 n^{4/3}} \left(\frac{1}{\tau^4}\right),
\label{deltamax}\\
\frac{\Delta}{2\pi}&\approx5\frac{\delta_{\rm max}}{2\pi}
\approx\frac{6.63}{\alpha^4}\frac{\mu^3 a^2}{\hbar^3 n^{4/3}} \left(\frac{1}{\tau^4}\right).
\label{DeltaD}
\end{align}
It is interesting noting that for $\tau\gg\tau_c$ the linewidth $\Delta\sim1/\tau^4$ rapidly decreases
as a function of the pulse length. 
In contrast, for shorter pulses, i.e., for $h/E_b\ll\tau\ll\tau_c$, the dissociation probability is
drastically, reduced and with lineshape parameters given by: 
\begin{eqnarray}
\frac{\delta_{\rm max}}{2\pi}\approx\frac{2}{\tau}\left(\frac{3}{5}\right)^{1/2}~~\mbox{and}~~~
\frac{\Delta}{2\pi}\approx\frac{1}{4}\left(\frac{5}{3}\right)^{1/2}\frac{\delta_{\rm max}}{2\pi}\label{Short}.
\end{eqnarray}
Therefore, for short pulses, since $\Delta/2\pi\sim1/\tau$, one would expect broad lineshapes and, consequently, substantially 
more heating than for long pulses. We note that both our results for long and short pulses lead to 
$\Delta/2\pi\sim\delta_{\rm max}/2\pi$, which is in agreement with the experimental findings in Refs.~\cite{greiner2005prl}.
We also note that although our longest ($\tau=250$ms) and shortest ($\tau=25$ms) pulse lengths are 
not strongly in the $\tau\gg\tau_c$ and $\tau\ll\tau_c$ regimes, we still obtain a reasonable agreement 
between our numerical results for $\delta_{\rm max}/2\pi$ and $\Delta/2\pi$ and the ones from 
Eqs.~(\ref{deltamax})-(\ref{Short}).

Based on this analysis, it is clear that the conditions for efficient dissociation relies on the pulse length
as well as the time scale for molecular losses. These conditions can be expressed as
\begin{align}
\frac{\tau_c}{\tau}&\approx\frac{1.11}{\alpha^{4/3}}\left(\frac{\mu\:a^{2/3}}{\hbar\:n^{4/9}}\right)\frac{1}{\tau}\ll1,\label{PulseC}\\
\frac{\gamma\tau}{2}&\approx\frac{\tau}{2}\left(n \tilde\gamma \frac{\hbar a}{\mu}\right)\ll1\label{LossTauC},
\end{align}
For instance, from the above equations we can see that although increasing the density improves the
condition for long pulses [Eq.~(\ref{PulseC})] it can lead to stronger losses [Eq.~(\ref{LossTauC})]. 
In fact, based on the different dependence on the experimentally relevant parameters in Eqs.~(\ref{PulseC})
and (\ref{LossTauC}) one can draw general conclusions concerning dissociation efficiency in different 
regimes. For high densities, for instance, Eqs.~(\ref{PulseC}) and (\ref{LossTauC}) indicate that efficient dissociation 
can only be achieved for small values of $a$, in order to minimize loss effects. 
On the other hand,  Eqs.~(\ref{PulseC}) and (\ref{LossTauC}) also indicates that dissociation of very weakly bound 
Feshbach molecules (large $a$) can only be efficient if one now considers the regime of both low density and long pulses.
We note that, differently than association, increasing $a$ could in principle lead to a broader linewidth $(\Delta/2\pi\sim a^2)$ 
[see Eq.~(\ref{DeltaD})], resulting in dissociated atoms with higher kinetic energy $\hbar\Delta$ \cite{greiner2005prl}. 
However, due to the strong dependency of 
$\Delta/2\pi$ on $\tau$, it turned out to be much easier to obtain narrow linewidths for dissociation than association.

As mentioned in Section \ref{Theory} ---the discussion following Eq.~(\ref{PK})--- thermal effects in molecular dissociation can only be introduced
via the Doppler effect. For the parameters relevant to our problem, however, doppler-broadening, 
$\Delta \omega=\omega(k_BT/mc^2)^{1/2}$, is found to be negligible. For instance, for $^{87}$Rb the resonant frequency $\omega/2\pi$
for the transition $|10\rangle$-$|11\rangle$ 
is less than $30$ MHz for fields below 40 Gauss which corresponds to Doppler widths below 0.05 mHz at $T=1$nK.
Nevertheless, for dissociation one could expect very narrow linewidths for long pulses ($\tau\gg\tau_c$), the 
Doppler-broadening will, at some point, be the main factor determining the linewidth for molecular dissociation. 
Similar to the case of association, molecular dissociation could also be sensitive to mean-field shifts (estimated above to be
of the order of 0.1Hz). In the regime of long dissociation pulses, $\tau\gg\tau_c$, leading to very narrow linewidths (see Fig.~\ref{RF_Dissociation}),
mean-field shifts can in fact become important in determining the value of the detuning in which dissociation is maximum. 
However, in order to more precisely determine these mean-field shifts one would need to explore in details the 
nature of the molecule-molecule interactions and their universal properties \cite{dincao2009prl,dincao2009pra}, 
a task beyond the scope of our present study.

%%%%%%%%%%%%%%%%%%%%%%%%%%%%%%%
\section{Summary}

We have developed a simple theoretical model capable of describing association and dissociation of
weakly bound heteronuclear Feshbach molecules with oscillating, state-changing fields. Our model is nonpertubative 
and accounts for coherent effects such as Rabi-oscillations as well as incoherent phenomena associated with atomic 
and molecular losses. 
Our analysis shows that the ultralow temperature and density regimes expected on CAL are 
beneficial for studies of association and dissociation of Feshbach molecules as well as the coherent properties
of such processes.
Hence, not only is the typical utility of Feshbach molecular physics enhanced in space, but new applications also emerge. 
Notably, heteronuclear Feshbach molecules can be used to achieve exquisite control over the initial density and momentum 
states of dual-species atomic and molecular gases for space-based fundamental physics research \cite{Williams}. 
We note that, in most of our calculations, the effects of the losses are suppressed due to the low-density regime accessible on CAL. 
Nevertheless, it would be interesting to explore experimentally the regime in which losses are important 
\cite{weber2008pra} in order to observe possible shifts on the association/dissociation linewidth due to losses, as 
predicted by Eqs.~(\ref{Pm}) and (\ref{PK}). 

\acknowledgments

This research was carried out under a contract with the National Aeronautics and Space Administration. 
MK acknowledges support by the German space agency with funds provided
by the Federal Ministry of Economics and Technology under grant numbers
50WP1432, 50WM1132 and 1237.
JPD also acknowledges partial support from the U. S. National Science Foundation, grant number PHY-1307380.


\begin{thebibliography}{99}

\bibitem{chin2010rmp} C. Chin, R. Grimm, P. S. Julienne, and E. Tiesinga,
Rev. Mod. Phys. {\bf 82}, 1225 (2010).

\bibitem{koehler2006rmp} T. K\"ohler, Krzysztof G\'oral, and P. S. Julienne,
Rev. Mod. Phys. {\bf 78}, 1311 (2006).

%BECBCS-Studies

\bibitem{greiner2003Nature} M. Greiner, C. A. Regal, and D. Jin, Nature (London) {\bf 426}, 537 (2003).

\bibitem{cubizolles2003prl} J. Cubizolles, T. Bourdel, S. J. J. M. F. Kokkelmans, G. V. Shlyapnikov, and C. Salomon, Phys. Rev. Lett. {\bf 91}, 240401 (2003).

\bibitem{regal2004prla} C. A. Regal, M. Greiner, and D. S. Jin, Phys. Rev. Lett. {\bf 92}, 083201 (2004).

\bibitem{jochim2003Science} S. Jochim, M. Bartenstein, A. Altmeyer, G. Hendl, S. Riedl, C. Chin, J. H. Denschlag, R. Grimm, Science {\bf 302} 2101 (2003).

\bibitem{zwierlien2004prl} M. W. Zwierlein, C. A. Stan, C. H. Schunck, S. M. F. Raupach, A. J. Kerman, and W. Ketterle, Phys. Rev. Lett. {\bf 92}, 120403 (2004).

\bibitem{strecker2003prl} K. E. Strecker, G. B. Partridge, and R. G. Hulet,  Phys. Rev. Lett. {\bf 91} 080406 (2003). 

\bibitem{regal2004prlb} C. A. Regal, M. Greiner, and D. S. Jin, Phys. Rev. Lett. {\bf 92}, 040403 (2004).

\bibitem{zwierlein2004prl} M. W. Zwierlein, C. A. Stan, C. H. Schunck, S. M. F. Raupach, A. J. Kerman, and W. Ketterle, 
Phys. Rev. Lett. {\bf 92}, 120403 (2004).

\bibitem{chin2005Science} C. Chin, M. Bartenstein, A. Altmeyer, S. Riedl, S. Jochim, J. H. Denschlag, and R. Grimm, 
Science {\bf 305}, 1128 (2004).  

\bibitem{zwierlein2005Nature} M. W. Zwierlein, J. R. Abo-Shaeer, A. Schirotzek, C. H. Schunck and 
W. Ketterle, Nature (London) {\bf 435}, 1047 (2005).

%Ground state molecules ...

%Efficient state transfer in an ultracold dense gas of heteronuclear molecules
\bibitem{ospelkaus2008NatPhys} S. Ospelkaus, A. Pe'er, K.-K. Ni, J. J. Zirbel, B. Neyenhuis, S. Kotochigova, P. S. Julienne, J. Ye and D. S. Jin,
Nat. Phys. {\bf 4}, 622 (2008).

%A High Phase-Space-Density Gas of Polar Molecules
\bibitem{ni2008Science} K.-K. Ni, S. Ospelkaus, M. H. G. de Miranda, A. Pe'er, B. Neyenhuis, J. J. Zirbel1, S. Kotochigova, P. S. Julienne, D. S. Jin and J. Ye,
Science {\bf 322}, 231 (2008).

%Collisional Stability of Fermionic Feshbach Molecules
\bibitem{zirbel2008prl} J. J. Zirbel, K.-K. Ni, S. Ospelkaus, J. P. D'Incao, C. E. Wieman, J. Ye, and D. S. Jin,
Phys. Rev. Lett. {\bf 100}, 143201 (2008).

%Ultracold Fermionic Feshbach Molecules of Na23K40
\bibitem{wu2012prl} C.-H. Wu, J. W. Park, P. Ahmadi, S. Will, and M. W. Zwierlein,
Phys. Rev. Lett. {\bf 109}, 085301 (2012).

%Formation of ultracold fermionic NaLi Feshbach molecules
\bibitem{heo2012pra} M.-S. Heo, T. T. Wang, C. A. Christensen, T. M. Rvachov, D. A. Cotta, J.-H. Choi, Y.-R. Lee, and W. Ketterle,
Phys. Rev. A {\bf 86}, 021602(R) (2012).

%Ultracold mixtures of atomic 6Li and 133Cs with tunable interactions
\bibitem{tung2013pra} S.-K. Tung, C. Parker, J. Johansen, C. Chin, Y. Wang, and P. S. Julienne,
Phys. Rev. A {\bf 87}, 010702(R) (2013).

%Observation of interspecies 6Li-133Cs Feshbach resonances
\bibitem{repp2013pra} M. Repp, R. Pires, J. Ulmanis, R. Heck, E. D. Kuhnle, M. Weidem\"uller, and E. Tiemann
Phys. Rev. A {\bf 87}, 010701(R) (2013).

%Production of optically trapped RbCs87 Feshbach molecules
\bibitem{koppinger2014pra} M. P. K\"oppinger, D. J. McCarron, D. L. Jenkin, P. K. Molony, H.-W. Cho, S. L. Cornish, C. R. LeSueur, 
C. L. Blackley, and J. M. Hutson, Phys. Rev. A {\bf 89}, 033604 (2014).

%Observation of Feshbach resonances between ultracold Na and Rb atoms
\bibitem{wang2013pra} F. Wang, D. Xiong, X. Li, D. Wang, and E. Tiemann, 
Phys. Rev. A {\bf 87}, 050702(R) (2013).

%Towards the production of ultracold ground-state RbCs molecules: Feshbach resonances, weakly bound states, and the coupled-channel model
\bibitem{takekoshi2012pra} T. Takekoshi, M. Debatin, R. Rameshan, F. Ferlaino, R. Grimm, H.-C. N\"agerl, C. R. LeSueur, J. M. Hutson, P. S. Julienne, S. Kotochigova, 
and E. Tiemann, Phys. Rev. A {\bf 85}, 032506 (2012).

%Giant Feshbach resonances in Li6-Rb85 mixtures
\bibitem{deh2010pra} B. Deh, W. Gunton, B. G. Klappauf, Z. Li, M. Semczuk, J. Van Dongen, and K. W. Madison,
Phys. Rev. A {\bf 82}, 020701(R) (2010).

% Efimov physics

% Few-body review
\bibitem{braaten2006PR} E. Braaten and H.-W. Hammer, Phys. Rep. {\bf 428}, 259 (2006).

\bibitem{wang2013AAMOP} Y. Wang, J. P. D'Incao, and B. D. Esry,  Adv. At. Mol. Opt. Phys. {\bf 62}, 1 (2013).

%Observation of Heteronuclear Atomic Efimov Resonances
\bibitem{barontini2009prl} G. Barontini, C. Weber, F. Rabatti, J. Catani, G. Thalhammer, M. Inguscio, and F. Minardi,
Phys. Rev. Lett. {\bf 103}, 043201 (2009).

%Tests of Universal Three-Body Physics in an Ultracold Bose-Fermi Mixture
\bibitem{bloom2013prl} R. S. Bloom, M.-G. Hu, T. D. Cumby, and D. S. Jin,
Phys. Rev. Lett. {\bf 111}, 105301 (2013).

%Geometric Scaling of Efimov States in a Li6?Cs133 Mixture
\bibitem{tung2014prl} S.-K. Tung, K. Jim\'enez-Garc\'ia, J. Johansen, C. V. Parker, and C. Chin,
Phys. Rev. Lett. {\bf 113}, 240402 (2014).

%Observation of Efimov Resonances in a Mixture with Extreme Mass Imbalance
\bibitem{pires2014prl} R. Pires, J. Ulmanis, S. H\"afner, M. Repp, A. Arias, E. D. Kuhnle, and M. Weidem\"uller,
Phys. Rev. Lett. {\bf 112}, 250404 (2014).

%Efimov Resonance and Three-Body Parameter in a Lithium-Rubidium Mixture
\bibitem{maier20a5prl} R. A. W. Maier, M. Eisele, E. Tiemann, and C. Zimmermann,
Phys. Rev. Lett. {\bf 115}, 043201 (2015).


%Detection of motional entanglement

%Probing Pair-Correlated Fermionic Atoms through Correlations in Atom Shot Noise
\bibitem{greiner2005prl} M. Greiner, C. A. Regal, J. T. Stewart, and D. S. Jin,
Phys. Rev. Lett. {\bf 94}, 110401 (2005).

%Quantum states of Bose-Einstein condensates formed by molecular dissociation
\bibitem{poulsen2001pra} U. V. Poulsen and K. M{\o}lmer,
Phys. Rev. A {\bf 63}, 023604 (2001).

%Quantum correlated twin atomic beams via photodissociation of a molecular Bose-Einstein condensate
\bibitem{kherunssyan2002pra} K. V. Kheruntsyan and P. D. Drummond,
Phys. Rev. A {\bf 66}, 031602(R) (2002).

%Matter-wave amplification and phase conjugation via stimulated dissociation of a molecular Bose-Einstein condensate
\bibitem{kheruntsyan2005pra} K. V. Kheruntsyan,
Phys. Rev. A {\bf 71}, 053609 (2005).

%Formation of a molecular Bose-Einstein condensate and an entangled atomic gas by Feshbach resonance
\bibitem{yurovsky2003pra} V. A. Yurovsky and A. Ben-Reuven,
Phys. Rev. A {\bf 67}, 043611 (2003).

%Spatial Pair Correlations of Atoms in Molecular Dissociation
\bibitem{savage2007prl} C. M. Savage and K. V. Kheruntsyan,
Phys. Rev. Lett. {\bf 99}, 220404 (2007).

%Einstein-Podolsky-Rosen Correlations via Dissociation of a Molecular Bose-Einstein Condensate
\bibitem{kheruntsyan2005prl} K. V. Kheruntsyan, M. K. Olsen, and P. D. Drummond,
Phys. Rev. Lett. {\bf 95}, 150405 (2005).

%High-fidelity entanglement via molecular dissociation in integrated atom optics
\bibitem{zhao2007pra} B. Zhao, Z.-B. Chen, J.-W. Pan, J. Schmiedmayer, A. Recati, G. E. Astrakharchik, and T. Calarco,
Phys. Rev. A {\bf 75}, 042312 (2007).

%Pairing mean-field theory for the dynamics of dissociation of molecular Bose-Einstein condensates
\bibitem{davis2008pra} M. J. Davis, S. J. Thwaite, M. K. Olsen, and K. V. Kheruntsyan,
Phys. Rev. A {\bf 77}, 023617 (2008).

%Bell Test for the Free Motion of Material Particles
\bibitem{gneiting2008prl} C. Gneiting and K. Hornberger,
Phys. Rev. Lett. {\bf 101}, 260503 (2008).

%Molecular Feshbach dissociation as a source for motionally entangled atoms
\bibitem{gneiring2010pra} C. Gneiting and K. Hornberger,
Phys. Rev. A {\bf 81}, 013423 (2010).

%Precision measurement (fundamental constants)

%Enhanced sensitivity to fundamental constants in ultracold atomic and molecular systems near feshbach resonances
\bibitem{chin2006prl}C. Chin and V. V. Flambaum, 
Phys. Rev. Lett. {\bf 96}, 230801 (2006).

%Ultracold molecules: new probes on the variation of fundamental constants
\bibitem{chin2009njp} C. Chin, V. V. Flambaum, and M. G. Kozlov,
New J. Phys. {\bf 11}, 055048 (2009).

%Sensitivity of ultracold-atom scattering experiments to variation of the ne-structure constant.
\bibitem{borschevsky2011pra} A. Borschevsky, K. Beloy, V. V. Flambaum, and P. Schwerdtfeger, 
Phys. Rev. A {\bf 83}, 052706 (2011).

%Photoassociation of ultracold molecules near a feshbach resonance as a probe of the electron-proton mass ratio variation.
\bibitem{gacesa2014jms} M. Gacesa and R. C\^ot\'e,
J. Mol. Spectrosc. {\bf 300}, 124 (2014).

% Interest in space research

\bibitem{Decadal}
Committee for the Decadal Survey on Biological and Physical Sciences in Space; National Research Council. \emph{Recapturing a Future for Space Exploration: Life and Physical Sciences Research for a New Era}. The National Academies Press (2011).

\bibitem{Turyshev2008}
S. G. Turyshev, U. E. Israelsson, M. Shao, N. Yu, A. Kusenko, E. L. Wright, C. W. F. Everitt, M. Kasevich, J. A. Lipa, J. C. Mester, R. D. Reasenberg, R. L. Walsworth, N. Ashby, H. Gould, and H. J. Paik, Int. J. Mod. Phys. D {\bf 16}, 1879 (2007).

\bibitem{Binns2009}
D.A. Binns, N. Randoa, and L. Cacciapuoti, Adv. Space Res. {\bf 43}, 1158 (2009).

\bibitem{Muntinga2013PRL}
H. Muntinga, H. Ahlers, M. Krutzik, A. Wenzlawski, S. Arnold, D. Becker, K. Bongs, H. Dittus, H. Duncker, N. Gaaloul, C. Gherasim, E. Giese, C. Grzeschik, T. W. Hansch, O. Hellmig, W. Herr, S. Herrmann, E. Kajari, S. Kleinert, C. Lammerzahl, W. Lewoczko-Adamczyk, J. Malcolm, N. Meyer, R. Nolte, A. Peters, M. Popp, J. Reichel, A. Roura, J. Rudolph, M. Schiemangk, M. Schneider, S. T. Seidel, K. Sengstock, V. Tamma, T. Valenzuela, A. Vogel, R. Walser, T. Wendrich, P. Windpassinger, W. Zeller, T. van Zoest, W. Ertmer, W. P. Schleich, and E. M. Rasel, Phys. Rev. Lett. {\bf 110}, 093602 (2013).

\bibitem{Zoest2010Science}
T. van Zoest, N. Gaaloul, Y. Singh, H. Ahlers, W. Herr, S. T. Seidel, W. Ertmer, E. Rasel, M. Eckart, E. Kajari, S. Arnold, G. Nandi, W. P. Schleich, R. Walser, A. Vogel, K. Sengstock, K. Bongs, W. Lewoczko-Adamczyk, M. Schiemangk, T. Schuldt, A. Peters, T. Konemann, H. Muntinga, C. Lammerzahl, H. Dittus, T. Steinmetz, T. W. Hansch, and J. Reichel, Science, {\bf 328}, 1540 (2010).

\bibitem{Stern2009EPJD}
G. Stern, B. Battelier, R. Geiger, G. Varoquaux, A. Villing, F. Moron, O. Carraz, N. Zahzam, Y. Bidel, W. Chaibi, F. P. Dos Santos, A. Bresson, A. Landragin and P. Bouyer, Eur. Phys. J. D {\bf 53}, 353 (2009).

\bibitem{Yu2006APB}
N. Yu, J. M. Kohel, J. R. Kellogg, and L. Maleki, Applied Physics B {\bf 84}, 647 (2006).

%\bibitem{13NPJ}
%G. M. Tino , F. Sorrentino , D. Aguilera, B. Battelier , A. Bertoldi , Q. Bodart , K. Bongs, P. Bouyer , C. Braxmaier, L. Cacciapuoti , N. Gaaloul , N. Gurlebeck, M. Hauth, S. Herrmann, M. Krutzik, A. Kubelka, A. Landragin , A. Milke, A. Peters, E. M. Rasel , E. Rocco, C. Schubert , T. Schuldt, K. Sengstock , A. Wicht, Nuclear Physics B (Proc. Suppl.), {\bf 243-244}, 203 (2013).

\bibitem{STEQUEST} T. Schuldt, C. Schubert, M. Krutzik {\em et al.}, Experimental Astronomy {\bf 39}, 167 (2015);
D. N. Aguilera, H. Ahlers, B. Battelier, {\em et al.}, Classical and Quantum Gravity {\bf 31}, 115010 (2014).

\bibitem{Williams2016NJP}
J. R. Williams, S w. Chiow, H. Mueller, and N. Yu,  New J. Phys., {\bf 18}, 025018 (2016).

\bibitem{Chu1986}
S. Chu, J. E. Bjorkholm, A. Ashkin, J. P. Gordon, and L. W. Hollberg, Opt. Lett. {\bf 11}, 73 (1986).

\bibitem{Ammann1997PRL}
H. Ammann and N. Christensen, Phys. Rev. Lett. {\bf 78}, 2088 (1997).

\bibitem{Myrskog2000PRA}
S. H. Myrskog, J. K. Fox, H. S. Moon, J. B. Kim, and A. M. Steinberg, Phys. Rev. A {\bf 61} 053412 (2000).

\bibitem{Leanhardt2003Science}
A.E. Leanhardt, T.A. Pasquini, M. Saba, A. Schirotzek, Y. Shin, D. Kielpinski, D.E. Pritchard, and W. Ketterle, Science {\bf 301}, 1513 (2003).

\bibitem{Hansen2013PRA}
A. H. Hansen, A. Y. Khramov, W. H. Dowd, A. O. Jamison, B. Plotkin-Swing, R. J. Roy, and S. Gupta, Phys. Rev. A, {\bf 87}, 013615 (2013).

\bibitem{Davis2002}
M. J. Davis and C. W. Gardiner, J. Phys. B: At. Mol. Opt. Phys. {\bf 35}, 733 (2002).

\bibitem{Thompson2013}
R. J. Thompson, Science Envelope Requirements Document (SERD) for Cold Atom Laboratory, JPL Technical Report (2013).

% KRb experiments

\bibitem{klempt2008pra}
C. Klempt, T. Henninger, O. Topic, M. Scherer, L. Kattner, E. Tiemann, W. Ertmer, and J. J. Arlt,
Phys. Rev. A {\bf 78}, 061602R (2008).

\bibitem{weber2008pra}
C. Weber, G. Barontini, J. Catani, G. Thalhammer, M. Inguscio, and F. Minardi,
Phys. Rev. A {\bf 78}, 061601(R) (2008).

%Double Species Bose-Einstein Condensate with Tunable Interspecies Interactions
%G. Thalhammer, G. Barontini, L. De Sarlo, J. Catani, F. Minardi, and M. Inguscio
%Phys. Rev. Lett. 100, 210402 ? Published 29 May 2008

%K40?Rb87 Feshbach resonances: Modeling the interatomic potential
\bibitem{klempt2007pra} C. Klempt, T. Henninger, O. Topic, J. Will, W. Ertmer, E. Tiemann, and J. Arlt,
Phys. Rev. A {\bf 76}, 020701(R) (2007).

%Near-threshold model for ultracold KRb dimers from interisotope Feshbach spectroscopy
\bibitem{simoni2008pra}A. Simoni, M. Zaccanti, C. D'Errico, M. Fattori, G. Roati, M. Inguscio, and G. Modugno,
Phys. Rev. A {\bf 77}, 052705 (2008).

%Collisional and molecular spectroscopy in an ultracold Bose?Bose mixture
\bibitem{thalhammer2009njp} G. Thalhammer, G. Barontini, J. Catani, F Rabatti, C. Weber, A. Simoni, F Minardi, and M, Inguscio,
New J. Phys. {\bf 11}, 055044 (2009).

% Floquet

\bibitem{chu2004pr} S.-I Chu and D. A. Telnov, Phys. Rep. {\bf 390}, 1 (2004).

% Zero-range

\bibitem{huang1957pr} K. Huang and C. N. Yang, Phys. Rev. {\bf 105}, 767 (1957).

% Fermi Energy

\bibitem{borca2003njp} B. Borca, D. Blume, and C. H. Greene, New J. Phys. {\bf 5}, 111 (2003).

\bibitem{goral2004jpb} K. G\'oral, T. K\"ohler, S. A. Gardiner, E. Tiesinga, and P. S. Julienne, J. Phys. B: At., Mol. Opt. Phys. {\bf 37}, 3457 (2004).

\bibitem{stecher2007prl} J. von Stecher and C. H. Greene,
Phys. Rev. Lett. {\bf 99}, 090402 (2007).

\bibitem{sykes2014pra} A. G. Sykes, J. P. Corson, J. P. D'Incao, A. P. Koller, C. H. Greene, A. M. Rey, K. R. A. Hazzard, and J. L. Bohn,
Phys. Rev. A {\bf 89}, 021601(R) (2014).

\bibitem{corson2015pra} J. P. Corson and J. L. Bohn,
Phys. Rev. A {\bf 91}, 013616 (2015).

% RWA

\bibitem{CohenTannoudji1992} 
C. Cohen-Tannoudji, J. Dupont-Roc, and G. Grynberg, Atom-Photon Interactions: Basic Processes and Applications (John Wiley \& Sons, New York, 1992).

% Association & Dissociation

\bibitem{chin2005pra} 
C. Chin and P. S. Julienne, Phys. Rev. A {\bf 71} 012713 (2005).

\bibitem{hanna2007pra} T. M. Hanna, T. K\"ohler, and K. Burnett, 
Phys. Rev. A {\bf 75}, 013606 (2007).

% Three-body losses

\bibitem{helfrich2010pra}
K. Helfrich, H.-W. Hammer, and D. S. Petrov, Phys. Rev. A {\bf 81}, 042715 (2010).

\bibitem{dincao2004prl}
J. P. D'Incao, H. Suno, and B. D. Esry, Phys. Rev. Lett. 93, 123201 (2004).

\bibitem{wang2012prlb}Y. Wang, J. Wang, J. P. D'Incao, and C. H. Greene,
Phys. Rev. Lett. {\bf 109}, 243201 (2012).

% Quantum Degeneracy issues

%Production Efficiency of Ultracold Feshbach Molecules in Bosonic and Fermionic Systems
\bibitem{hodby2005prl} E. Hodby, S. T. Thompson, C. A. Regal, M. Greiner, A. C. Wilson, D. S. Jin, E. A. Cornell, and C. E. Wieman,
Phys. Rev. Lett. {\bf 94}, 120402 (2005).

%Ultracold Molecule Production via a Resonant Oscillating Magnetic Field
\bibitem{thompson2005prl} S. T. Thompson, E. Hodby, and C. E. Wieman,
Phys. Rev. Lett. {\bf 95}, 190404 (2005).

%Observation of Heteronuclear Feshbach Molecules from a Rb85?Rb87 Gas
\bibitem{papp2006prl} S. B. Papp and C. E. Wieman,
Phys. Rev. Lett. {\bf 97}, 180404 (2006).

%Theory of Feshbach molecule formation in a dilute gas during a magnetic field ramp
\bibitem{williams2006njp} J. E. Williams, N. Nygaard, and C. W. Clark,
New J. Phys {\bf 8}, 150 (2006).

%Universal Four-Boson States in Ultracold Molecular Gases: Resonant Effects in Dimer-Dimer Collisions
\bibitem{dincao2009prl} J. P. D'Incao, J. von Stecher, and C. H. Greene,
Phys. Rev. Lett. {\bf 103}, 033004 (2009).

%Dimer-dimer collisions at finite energies in two-component Fermi gases
\bibitem{dincao2009pra} J. P. D'Incao, S. T. Rittenhouse, N. P. Mehta, and C. H. Greene,
Phys. Rev. A {\bf 79}, 030501(R) (2009).

\bibitem{Williams} M. Krutzik, E. Elliott, J. R. Williams and J. P. D'Incao, in preparation.

\end{thebibliography}
\end{document}